\def\mode{1} 
\pdfoutput=1
\if 0\mode
    \documentclass[preprint,superscriptaddress,floatfix]{revtex4-2}
    \usepackage[left]{lineno}
    \linenumbers
\else
    \documentclass[aps,pra,twocolumn,superscriptaddress,floatfix]{revtex4-2}
\fi

\usepackage{placeins}
\usepackage{afterpage}
\usepackage{environ}
\usepackage{fullpage}
\usepackage{amssymb,amsmath}
\usepackage[utf8x]{inputenc}
\usepackage[T1]{fontenc}
\usepackage{siunitx}
\usepackage[version=3]{mhchem}
\usepackage[dvipsnames]{xcolor}
\usepackage{xfrac}
\usepackage{tabularx}
\usepackage{multirow}
\usepackage[caption=false]{subfig} 

\newcommand\captionof[1]{\def\@captype{#1}\caption}

\usepackage[unicode=true]{hyperref}
\hypersetup{colorlinks=false,
            pdfborder={0 0 0}}
\usepackage{graphicx}
\makeatletter
\def \figwidth{
    \if 0\mode
        0.5\textwidth
    \else
        \columnwidth
    \fi
}
\makeatother

\makeatletter
\newwrite\remember@figures
\AtBeginDocument{\InputIfFileExists{\jobname.dft}{}{}\immediate\openout\remember@figures=\jobname.dft}
\AtEndDocument{\immediate\closeout\remember@figures}
\NewEnviron{dfigure}[1]{%
    \immediate\write\remember@figures{%
        \noexpand\rememberfigure{#1}{\unexpanded\expandafter{\BODY}}%
    }%
}
\NewEnviron{dfigure*}[1]{%
    \immediate\write\remember@figures{%
        \noexpand\rememberfiguretc{#1}{\unexpanded\expandafter{\BODY}}%
    }%
}
\newcommand{\placefigure}[2][tp]{%
    \csname remembered@figure@#2\endcsname{#1}%
}
\newcommand{\rememberfigure}[2]{%
    \global\@namedef{remembered@figure@#1}##1{%
        \begin{figure}[##1]#2\end{figure}%
    }%
}
\newcommand{\rememberfiguretc}[2]{%
    \global\@namedef{remembered@figure@#1}##1{%
        \begin{figure*}[##1]#2\end{figure*}%
    }%
}
\makeatother

\begin{document}
\title{Frustration- and doping-induced magnetism in a Fermi-Hubbard simulator}

\newcommand{\harvard}{Department of Physics, Harvard University, 17 Oxford St., Cambridge, MA 02138, USA}
\newcommand{\UCDavis}{Department of Physics, University of California, Davis, California 95616, USA}
\author{Muqing~Xu}
\author{Lev~Haldar~Kendrick}
\author{Anant~Kale}
\author{Youqi~Gang}
\author{Geoffrey~Ji}
\affiliation{\harvard}
\author{Richard~T.~Scalettar}
\affiliation{\UCDavis}
\author{Martin~Lebrat}
\author{Markus~Greiner}
\affiliation{\harvard}

\date{\today}

\begin{abstract}

Geometrical frustration in strongly correlated systems can give rise to a plethora of novel ordered states and intriguing magnetic phases such as quantum spin liquids \cite{anderson_resonating_1973, balents_spin_2010, zhou_quantum_2017}. Promising candidate materials for such phases \cite{williams_organic_1991, kino_phase_1996, shimizu_spin_2003} can be described by the Hubbard model on an anisotropic triangular lattice, a paradigmatic model capturing the interplay between strong correlations and magnetic frustration \cite{laubach_phase_2015, szasz_chiral_2020, motrunich_variational_2005, wietek_mott_2021, zhu_doped_2022}. However, the fate of frustrated magnetism in the presence of itinerant dopants remains unclear, as well as its connection to the doped phases of the square Hubbard model \cite{lee_doping_2006}. Here, we probe the local spin order of a Hubbard model with controllable frustration and doping, using ultracold fermions in anisotropic optical lattices continuously tunable from a square to a triangular geometry. At half-filling and strong interactions $U/t \approx 9$, we observe at the single-site level how frustration reduces the range of magnetic correlations and drives a transition from a collinear N\'eel antiferromagnet to a short-range correlated $120^\circ$ spiral phase. Away from half-filling, the triangular limit shows enhanced antiferromagnetic correlations on the hole-doped side,
and a reversal to ferromagnetic correlations at particle dopings above 20\%, hinting at the role of kinetic magnetism in frustrated systems. This work paves the way towards exploring possible chiral ordered or superconducting phases in triangular lattices \cite{szasz_chiral_2020, song_doping_2021}, and realizing $t-t'$ square lattice Hubbard models that may be essential to describe superconductivity in cuprate materials \cite{PhysRevLett.87.047003}.

\end{abstract}

\maketitle

\section*{Introduction}

\if 1\mode
\placefigure[!t]{f1}
\fi

The collective properties of spins with antiferromagnetic interactions crucially depend on the geometry of the lattice they inhabit \cite{wannier_antiferromagnetism_1950}. On a square lattice, spins form a N\'eel order with antialigned neighbours; in contrast, their mutual antiparallel alignment cannot be satisfied on a triangular lattice, which is the simplest model for geometric frustration and features non-trivial spin order. This frustrated spin order is associated with a massive ground state degeneracy with enhanced quantum fluctuations, and may lead to exotic phases of matter such as quantum spin liquids \cite{anderson_resonating_1973,lee_end_2008,balents_spin_2010,savary_quantum_2016}.

The Hubbard Hamiltonian is one of the most fundamental models describing the emergence of quantum magnetism among spin-$1/2$ electrons with kinetic energy $t$ and interaction energy $U$. On the non-frustrated square lattice, it is thought to capture the essential physics of the strongly-correlated electrons in the doped high-temperature superconducting cuprate materials \cite{lee_doping_2006}. Interpolating the Hubbard model between square and triangular lattices has important practical value to accurately describe a broader class of correlated materials with structural anisotropy, including layered organic compounds believed to host quantum spin liquid phases \cite{zhou_quantum_2017}. Anisotropic triangular Hubbard models would furthermore provide a minimal model to understand the competition between charge dopants and magnetism with frustration away from half-filling, where much less is known since numerical calculations are challenging due to the absence of particle-hole symmetry.

Ultracold fermions in optical lattices form a pristine realisation of the Hubbard model. They can be used for the quantum simulation of frustrated systems \cite{yang_site-resolved_2021,mongkolkiattichai_quantum_2022}, shedding light both on its half-filled and doped phases with site-resolved observables. In this work, we realize a Fermi-Hubbard system with tuneable frustration and investigate its magnetic order upon doping with single-site resolution in the regime of intermediate to strong interactions $U/t \approx 9$. 
We explicitly implement tuneable tunnelling anisotropy and investigate the combined effect of frustration and doping on magnetic order at temperatures $T/t \lesssim 0.4$ comparable to or lower than the spin exchange energy. This is in contrast to concurrent work probing nearest-neighbour antiferromagnetic correlations on an isotropic triangular lattice \cite{mongkolkiattichai_quantum_2022}, and to previous studies focusing on frustrated classical magnetism with ultracold bosons \cite{struck_quantum_2011}.

Our system relies on a lattice formed by the interference of two orthogonal retro-reflected laser beams whose relative phase is actively stabilised \cite{tarruell_creating_2012,sebby-strabley_preparing_2007}. With equal beam intensities, this interference realises a non-separable square lattice rotated by $45^\circ$. Tuneable frustration is introduced by an additional tunnelling term $t'$ along one diagonal of this square lattice (Fig.~\ref{fig:fig1}a) and controlled by the intensity imbalance between the two beams (Fig.~\ref{fig:fig1}b), in contrast to previous realisations of lattices with three beams and a $120^\circ$ rotational invariance \cite{struck_quantum_2011,jo_ultracold_2012,yamamoto_single-site-resolved_2020,yang_site-resolved_2021,trisnadi_design_2022}. As a result, our geometry can be smoothly changed from a square lattice for $t'/t = 0$ to an isotropic triangular lattice at $t'/t = 1$, and undergoes a dimensional crossover to weakly coupled 1D chains in the limit $t'/t \gg 1$.

\if 1\mode
\placefigure[!t]{f2}
\fi

\if 1\mode
\placefigure[!t]{f3}
\fi

We prepare a balanced mixture of fermionic $^6$Li atoms in the two lowest hyperfine states into this tuneable optical lattice by adiabatically ramping the lattice powers within $\SI{160}{\milli\second}$. We set the $s-$wave scattering length to values $a_{s}=358\,a_{0}-432\,a_{0}$ by tuning the magnetic bias field in the vicinity of the Feshbach resonance at $832 \text{G}$, where $a_{0}$ denotes the Bohr radius. The system is well described by a single-band Hubbard model with nearest neighbour tunnelling $t=\SI{355(11)}{\hertz}-\SI{426(21)}{\hertz}$ and a tuneable diagonal tunnelling $t'=\SI{9.5(4)}{Hz}-\SI{370(6)}{Hz}$. Due to the underlying harmonic confinement of the laser beams, atoms are subject to a trapping potential and display a spatially varying density $n$ (Fig.~\ref{fig:fig1}d; see Methods).

\section*{N\'eel to Spiral Order Transition}

In the strong coupling limit where $U$ is greater than the bandwidth, the Hubbard Hamiltonian at half-filling can be approximated by an antiferromagnetic Heisenberg model with anisotropic spin exchange couplings $J^{(\prime)} = 4 {t^{(\prime)}}^2 / U$.
This anisotropic spin model already features rich magnetic properties. In the bipartite square lattice $J' = 0$, the ground state is an antiferromagnetic N\'eel state \cite{hirsch_antiferromagnetism_1989}. In contrast, frustration in the isotropic triangular lattice gives rise to a $120^{\circ}$ spiral N\'eel order \cite{singh_three-sublattice_1992, huse_simple_1988, jolicoeur_spin-wave_1989, capriotti_long-range_1999}. Classical spin-wave theory predicts a transition between antiferromagnetic N\'eel order to an incommensurate spin spiral phase at $J'/J \geq 0.5$ which smoothly evolves into $120^{\circ}$ order at $J'/J = 1$ \cite{trumper_spin-wave_1999, merino_heisenberg_1999}. In the quantum spin-1/2 Heisenberg model, the location of the transition point is expected to be shifted above the classical value of $0.5$ due to quantum fluctuations but its exact location is still an open question \cite{weihong_phase_1999}.

To shed light on the magnetic properties of the anisotropic triangular Hubbard model at intermediate $U/t$, we form a large Mott insulator of about $500$ atoms by adjusting the local chemical potential at the centre of the trap to approximately reach half-filling (Fig.~\ref{fig:fig1}c). We measure the spin-spin correlation function
\begin{align}
    C_{\mathbf{d}}(\mathbf{r})&=\frac{1}{S^2}\left( \langle \hat{S}^{z}_{\mathbf{r}}\hat{S}^{z}_{\mathbf{r + d}}  \rangle - \langle \hat{S}^{z}_{\mathbf{r}}\rangle \langle \hat{S}^{z}_{\mathbf{r + d}} \rangle\right)
\end{align}
between any pair of sites located at positions $\mathbf{r}$ and $\mathbf{r} \pm \mathbf{d}$, as described in our previous work \cite{parsons_site-resolved_2016}. 
We average this correlator within the central insulating region of about 200 sites, 
where the chemical potential variation due to harmonic confinement is minimal.
In the square lattice $t'/t = 0.0265(3)$, we observe strong antiferromagnetic correlations decaying exponentially with distance (Fig.~\ref{fig:correlations}a, left), visible as a spatially averaged correlator $C_{\mathbf{d}}$ with a staggered sign that gradually fades out as a function of bond distance $\mathbf{d}$ in a logarithmic colour scale. 
Comparing the measured nearest-neighbour spin correlators to those obtained from Determinantal Quantum Monte Carlo (DQMC) simulations at half-filling gives a fitted temperature of $T/t=0.26(1)$. 
As lattice anisotropy $t'/t$ and frustration are increased, the growing superexchange coupling $J'$ along the diagonal $\mathbf{d}=(1, 1)$ favours anti-aligned spins, which competes with the N\'eel ordering which favours ferromagnetic correlation between sites on the same sublattice. 
As a result, we observe a suppression of the range of the spin-spin correlations. 
The correlator $C_{(1, 1)}$ is furthermore weakened for moderate anisotropies $t'/t = 0.57(3)$ before changing its sign (purple data points, Fig.~\ref{fig:correlations}b) \cite{chang_discriminating_2013}. 
In the configuration closest to the triangular geometry, $t'/t = 0.97(4)$, the three correlators to the nearest triangular neighbours $C_{(1, 0)}$ , $C_{(0, 1}$ and $C_{(1, 1)}$ are consistently isotropic, with a residual difference by about $10\%$ due to technical limitations (see Methods). 
We also observe positive next-nearest correlations $C_{(2, 1)}$ that reflect the effective hexagonal symmetry of the correlation function $C_{\mathbf{d}}$ that also show a sign change (red data points, Fig.~\ref{fig:correlations}b). We observe a slight temperature increase as anisotropy $t'/t$ is increased, to $T/t=0.39(4)$ in the triangular lattice, which also contributes to the suppression of spin correlations dominated by frustration (see Methods). This heating may be due to increased laser noise with increasing lattice intensity imbalance.

In solid-state systems, magnetic transitions can be observed through changes in the symmetry of the spin structure factor, which can be measured, for example, via neutron scattering. Here, we obtain the spin structure factor $S^{zz}(\textbf{q})$ from the Fourier transformation of the real-space spin correlation function (see Methods). Antiferromagnetic N\'eel order in the square lattice appears as a well-defined peak at quasi-momentum $(\pi, \pi)$, the $M$ symmetry point of the first Brillouin zone (Fig.~\ref{fig:correlations}c). 
As we increase $t'/t$ this peak becomes anisotropic, broadening along the $K-K'$ direction.
For the triangular lattice case, we observe two distinct peaks at the $K$ and $K'$ points of the hexagonal Brillouin zone, indicative of the $120^{\circ}$ spiral order.
The short-range character of this $120^\circ$ order is evident from a global reduction and a broadening of the spin structure factor peaks (Fig.~\ref{fig:correlations}d).

\if 1\mode
\placefigure[!t]{f4}
\fi

\section*{Particle-Hole Asymmetry}

Interactions between itinerant charge and magnetic moments can lead to rich collective quantum phases.
One paradigm is doping a N\'eel-ordered Mott insulator, where the interplay between the kinetic energy of the mobile dopants and strong correlations is believed to underlie the physics of cuprates.
In the square lattice Hubbard model, however, N\'eel antiferromagnetism
is made particularly robust at half-filling by Fermi surface nesting and the absence of geometric frustration, which may obscure competing orders.
Doping frustrated systems where intriguing phases already arise at half-filling may bring distinct new physics \cite{song_doping_2021, zhu_doped_2022}.
Anisotropic triangular lattices can be seen as the simplest lattice that frustrates colinear N\'eel order and breaks the particle-hole symmetry through a single diagonal next-nearest-neighbour tunnelling $t'$. 

We investigate the effect of doping by increasing the central lattice filling to $n = 1.6$; together with a slow variation of the chemical potential due to the lattice confinement, this allows us to probe short-range spin correlations over a large range of both particle and hole dopings $\delta = n-1$ in the local density approximation (see Methods). 
As expected from the particle-hole symmetry in the band structure (Fig.~\ref{fig:doping}a), we find the nearest-neighbour spin correlation $C_{(1,0)}$ in the square lattice remains antiferromagnetic and decays similarly upon hole- or particle-doping $\pm \delta$ (Fig.~\ref{fig:doping}b, bottom), with a residual asymmetry explained by deviations of the underlying confinement from a radially symmetric harmonic potential.
In contrast, we observe particle-hole-asymmetric magnetism in the nearly isotropic triangular lattice $t'/t = 0.97(4)$ (Fig.~\ref{fig:doping}b, top). There, at the same temperature, antiferromagnetic correlations survive for a wide range of hole doping, whereas they are strongly suppressed with particle doping. Surprisingly, we find the correlator $C_{(1,0)}$ even becomes significantly ferromagnetic above a certain particle doping $\delta \gtrsim 0.2$, as confirmed in spin correlation maps (Fig.~\ref{fig:doping}c).

Asymmetric spin correlations are to be expected for a Fermi liquid due to the particle-hole asymmetry of the triangular lattice band structure: the shape and topology of the non-interacting Fermi surface changes drastically upon increasing density and separates into two disconnected parts centred around the symmetry points $K$ and $K'$ close to full filling $n = 2$ (Fig.~\ref{fig:doping}a).
However, comparing to DQMC simulations we note that the experimentally observed sign reversal at large particle doping and sharp asymmetric suppression of spin correlation close to half-filling are absent in a non-interacting system (see Methods). The agreement with DQMC simulations at the experimental interaction strength $U/t = 9.2(5)$ suggests that this asymmetric interplay of particle and hole dopants with magnetism is unique to interacting systems (Fig.~\ref{fig:doping}b, grey lines). This asymmetry and strongly weakened correlations at particle dopings $\delta \approx +0.5$ are similarly observed in anisotropic triangular geometries $t'/t \geq 0.57$ (Fig.~\ref{fig:doping}d).

One possible mechanism for the particle-hole asymmetry in the stability of spin correlations can be understood by considering one dopant on a triangular plaquette in the superexchange energy $J(')=0$ limit \cite{tasaki_hubbard_1997, morera_high-temperature_2022} similar to Nagaoka effect \cite{nagaoka_ferromagnetism_1966}. The dopant could minimize its kinetic energy when different hopping paths interfere constructively, which is decided by the effective sign of the tunnellings $t(')$ and the surrounding spin configuration. With the sign convention of this work, a particle dopant has $t(') > 0$ and a spin triplet, ferromagnetic configuration allows the dopant to hop with constructive quantum interference. A hole dopant, by contrast, has an effective $t(') < 0$, thus prefers a spin singlet, antiferromagnetic configuration. However, at finite $J(')$ and low temperatures $T \lesssim J(')$ as in our experiment, how such kinetic frustration competes with magnetic orders still remains an open question. 

Particle-hole asymmetry is particularly apparent in spin correlations along the diagonal bonds $C_{(1, \pm1)}$ (Fig.~\ref{fig:doping_nnn}a). In a square lattice, both correlators are equal, particle-hole-symmetric, and show a reversal from positive to negative for dopings $|\delta| \gtrsim 0.2$ \cite{parsons_site-resolved_2016}. As the anisotropy $t'/t$ increases, the nature of $C_{(1, 1)}$ changes from a next-nearest-neighbour to a nearest-neighbour and its value smoothly interpolates to the particle-hole-asymmetric correlator $C_{(1,0)}$ in the triangular lattice (Fig.~\ref{fig:doping}b). The ferromagnetic character of this correlator upon particle doping is most pronounced at $t'/t \approx 0.5$ (Fig.~\ref{fig:doping_nnn}b) and we find quantitative agreement with DQMC simulations.
Interestingly, increasing the frustration parameter $t'/t$ has the opposite effect on the other diagonal correlator $C_{(1, -1)}$, which becomes antiferromagnetic upon particle doping (Fig.~\ref{fig:doping_nnn}a, bottom).

\section*{Discussion and Outlook}
Possible scenarios for the appearance of ferromagnetism in the Hubbard model have been identified at the mean-field level \cite{hirsch_two-dimensional_1985}, for single dopants \cite{nagaoka_ferromagnetism_1966} or at high temperature in frustrated systems \cite{morera_high-temperature_2022, haerter_kinetic_2005}, but a complete theoretical picture in our regime of temperatures $T \lesssim J$ and strong correlations is missing. The existence of a van Hove singularity in the non-interacting density of states of the triangular lattice at a density $n = 3/2$, together with weak ferromagnetic correlations observed in DQMC simulations even at small interaction $U/t = 4$ (see Methods) could suggest that density of states may play a crucial role \cite{hanisch_ferromagnetism_1995, martin_itinerant_2008, merino_ferromagnetism_2006, weber_magnetism_2006, lee_triangular_2022}. Experimentally, our findings might be related to recent observations in transition metal dichalcogenide moir\'e materials \cite{tang_simulation_2020}. Quantum gas microscope experiments could help elucidate the microscopic processes underlying doping-induced magnetism through the measurement of spin-spin-charge correlations \cite{koepsell_imaging_2019,garwood_site-resolved_2022} as well as momentum-resolved spectroscopy \cite{brown_angle-resolved_2020}.

Further experimental studies at interactions close to the metal-to-insulator transition would also help shed light on a conjectured spin-liquid phase with broken time-reversal symmetry \cite{szasz_chiral_2020, chen_quantum_2022}. Moreover, through the addition of a third superlattice beam \cite{tarruell_creating_2012}, our tuneable experimental platform allows for exploring extensions of the Hubbard model directly related to cuprate materials, such as the $t-t'$ model, which could help explain the emergence of superconducting phases upon doping \cite{PhysRevLett.87.047003}.

\if 0\mode
\placefigure[!t]{f1}
\fi

\if 0\mode
\placefigure[!t]{f2}
\fi

\if 0\mode
\placefigure[!t]{f3}
\fi

\if 0\mode
\placefigure[!t]{f4}
\fi

\begin{dfigure*}{f1}
    \centering
    \noindent
    \includegraphics[width=\linewidth]{"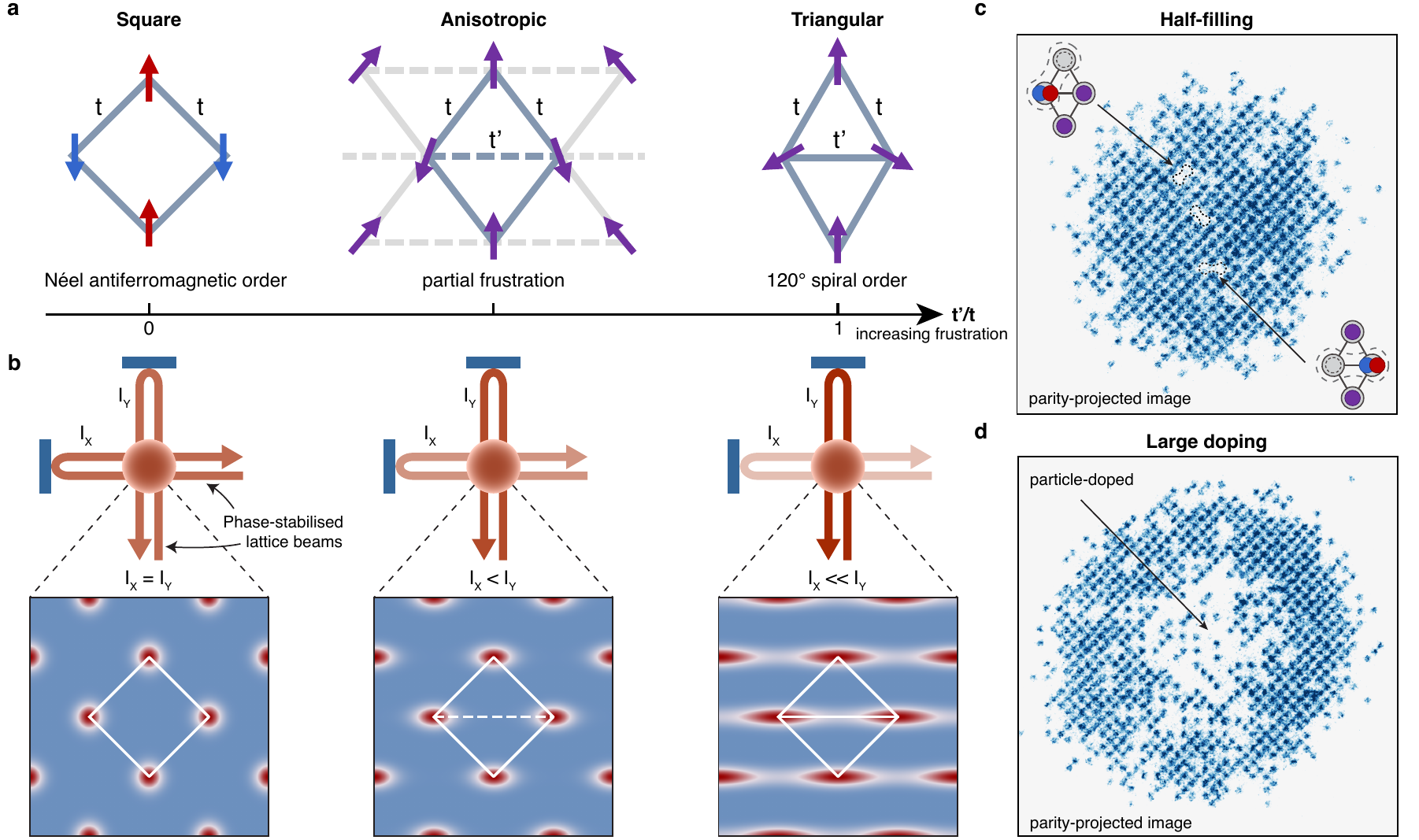"}
    \caption{\textbf{Probing frustration on a tuneable triangular lattice with a quantum gas microscope.}
    (\textbf{a}) A square lattice with coupling $t$ can be continuously transformed into a triangular lattice with an additional tuneable coupling $t'$ along one diagonal. Frustration is parameterized by the anisotropy ratio $t'/t$ and leads to a change in the magnetic ground state from an antiferromagnetic N\'eel order in the square lattice ($t'/t = 0$) to a $120^{\circ}$ spiral order in the isotropic triangular lattice ($t'/t = 1$), both in the classical and quantum Heisenberg limits.
    (\textbf{b}) We implement this tuneable lattice with two orthogonal retro-reflected lattice beams actively phase-locked to each other \cite{tarruell_creating_2012}. Their interference results in a non-separable square lattice potential rotated by $45^\circ$. Adjusting the intensity balance between the lattice beams reduces the potential barrier between a pair of diagonal neighbours and enhances tunnelling.
    (\textbf{c}) We realize a frustrated Fermi-Hubbard magnet by preparing a Mott insulator of about $500$ fermionic atoms in the tuneable lattice. Doublon-hole pairs appear due to quantum fluctuations at our finite interaction energy $U/t \approx 9$ and are imaged as pairs of empty sites connected by tunnellings $t(') > 0$ in a typical fluorescence picture due to parity-projected imaging (top, with anisotropy $t'/t = 0.26(1)$). 
    (\textbf{d}) Atomic density varies across the sample due to the presence of a radial confining potential, allowing us to locally investigate the effect of doping on magnetic order. Upon increasing the total atom number, a particle-doped region is imaged as a lighter disk inside a half-filled ring.
    }
    \label{fig:fig1}
\end{dfigure*}

\begin{dfigure*}{f2}
    \centering
    \noindent
    \includegraphics[width=\linewidth]{"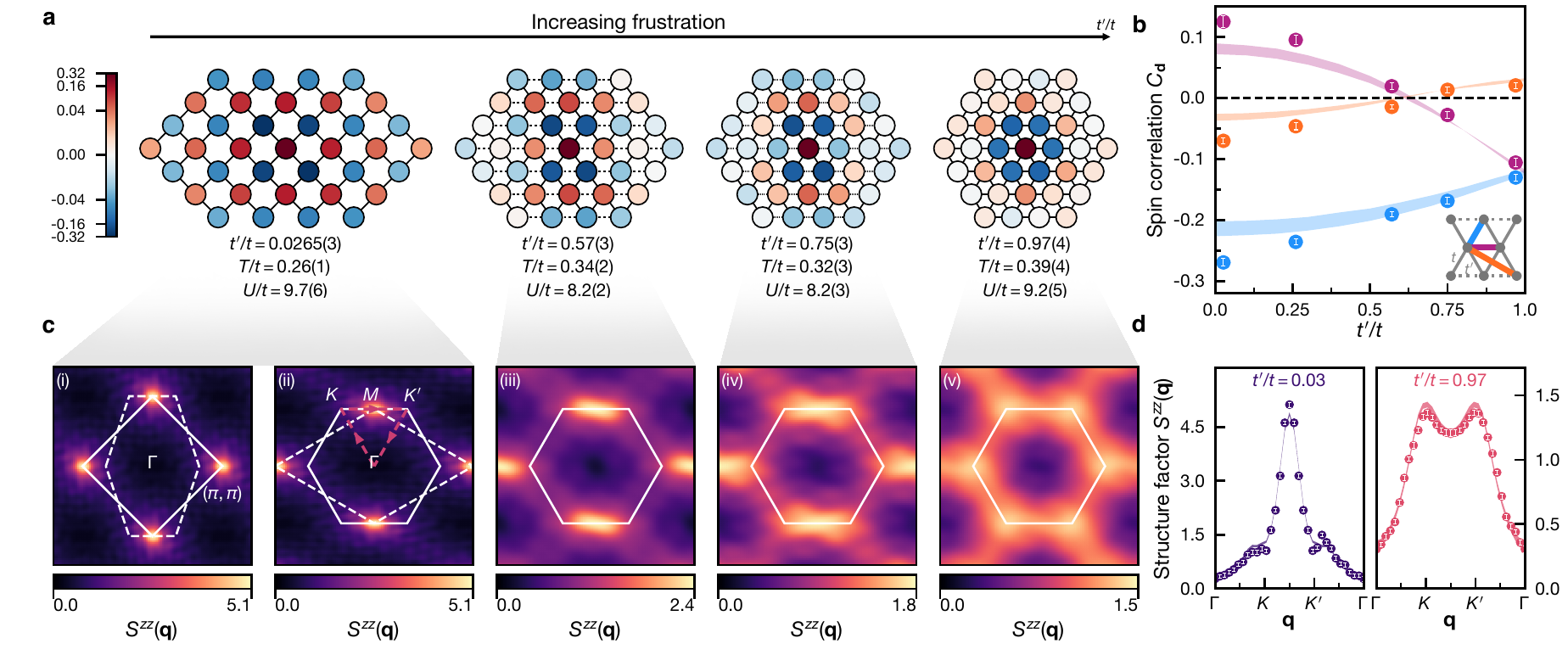"}
    \caption{\textbf{Frustrating short-range antiferromagnetic order in the square-to-triangular lattice transition.}
    (\textbf{a}) 
    Real-space spin correlation $C_{\mathbf{d}}$ as a function of displacement $\mathbf{d}$ averaged over the region at half-filling, plotted for increasing values of $t'/t$. The magnitude of antiferromagnetic correlations decreases with increasing frustration, parameterized by $t'/t$, and the symmetry of the correlation function changes from four-fold $D_4$ to six-fold $D_6$, i.e. hexagonal. Each panel is averaged over about $200$ sites and more than $400$ experimental realizations, with a typical standard error of $0.005$ (see Methods). The correlation plot grid is smoothly stretched horizontally to emphasize the change in connectivity.
    (\textbf{b}) Nearest-neighbour spin correlations across the $t$-bonds $C_{(1, 0)}$, across the $t'$-bonds $C_{(1, 1)}$, and next-nearest-neighbour correlation $C_{(2, 1)}$. $C_{(1, 0)}$ is decreasing with increasing frustration. $C_{(1, 1)}$ and $C_{(2, 1)}$ reverses its sign as diagonal neighbours in the square lattice with aligned spins become nearest neighbours in the triangular lattice with antialigned spins. Shaded bands: DQMC simulations at $U/t = 9.5$ and $T/t = 0.35-0.4$. Experimental temperatures are lower than the DQMC data here for $t'/t=0.0265(3)$ and $0.26(1)$.
    (\textbf{c}) 
    Measured spin structure factor $S^{zz}(\mathbf{q})$ plotted over the extended Brillouin zone (BZ) of the square lattice (i) and triangular lattice (ii-v). (i) and (ii) Antiferromagnetic order on the square lattice shows up as a single peak at quasi-momenta $(\pi, \pi)$ in the square BZ and a peak at the $M$ point of the hexagonal BZ. (ii-v) The single peak at the $M$ point broadens with increasing $t'/t$ and splits into two separate peaks at the $K$ and $K'$ points for $t'/t=0.97(4)$. The broad peaks in panel (v) indicate short-range $120^{\circ}$ order in the triangular lattice.
    (\textbf{d}) Cut of the spin structure factor $S^{zz}(\mathbf{q})$ along the $\Gamma-K-K'-\Gamma$ line (illustrated in (\textbf{c}), panel (ii)), showing the splitting of the spin structure factor peak between the square and triangular geometries. Shaded bands: DQMC simulations at $U/t=9.7, T/t=0.26$ (square) and $U/t=9.2,T/t=0.39$ (triangle), with widths propagated from experimental uncertainties (see Methods). The errorbars denote one s.e.m. and the number of repetitions can be found in Methods.}
    
    \label{fig:correlations}
\end{dfigure*}

\begin{dfigure*}{f3}
    \noindent
    \includegraphics[width=\linewidth]{"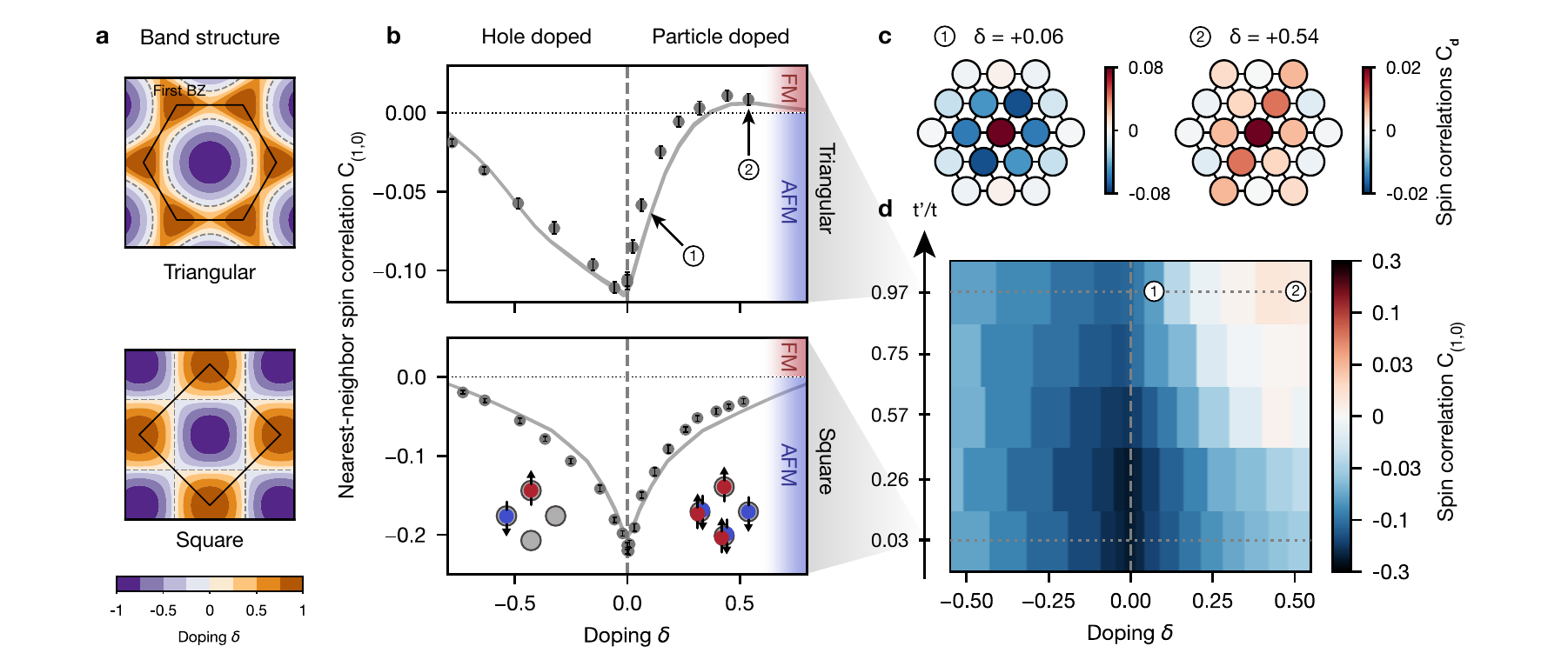"}
    \caption{\textbf{Particle-hole asymmetry of magnetic correlations and particle-doping-induced ferromagnetism.}
    (\textbf{a}) Particle-hole symmetry is broken in the triangular lattice, which is visible as a change of the shape and topology of the non-interacting band structure above (purple) and below half-filling (brown). In the particle-hole-symmetric square lattice, antiferromagnetic order is enhanced at half-filling due to a strongly nested square Fermi surface (dashed lines) for weak interactions.
    (\textbf{b}) Doping leads to a decrease of the magnitude of the antiferromagnetic spin correlations averaged over the $t$ lattice bonds $C_{(1, 0)}$ and $C_{(0, 1)}$ (grey points) compared to the half-filling value (vertical dashed line). A pronounced particle-hole asymmetry emerges in the triangular lattice, while the correlation function is symmetric in the square lattice. Grey lines show DQMC simulations at $U/t = 9.1$, $T/t = 0.44$ (triangular) and $U/t = 9.0$ and $T/t = 0.40$ (square).
    (\textbf{c}) Spin correlation maps in the triangular lattice at dopings $\delta = 0.06$ and $\delta = 0.54$. Nearest-neighbour correlations change to weakly ferromagnetic at $\delta = 0.54$ with values $C_{(1, 0)} = 0.011(6)$ and $C_{(1, 1)} = 0.006(5)$.
    (\textbf{d}) Nearest-neighbour spin correlations over the $t$ lattice bonds as a function of doping $\delta$ and lattice anisotropy $t'/t$, showing a large region of weakened correlations at positive doping and $t'/t > 0.5$. The errorbars denote one s.e.m. and the number of repetitions can be found in Methods.}
    \label{fig:doping}
\end{dfigure*}

\begin{dfigure*}{f4}
    \noindent
    \includegraphics[width=\linewidth]{"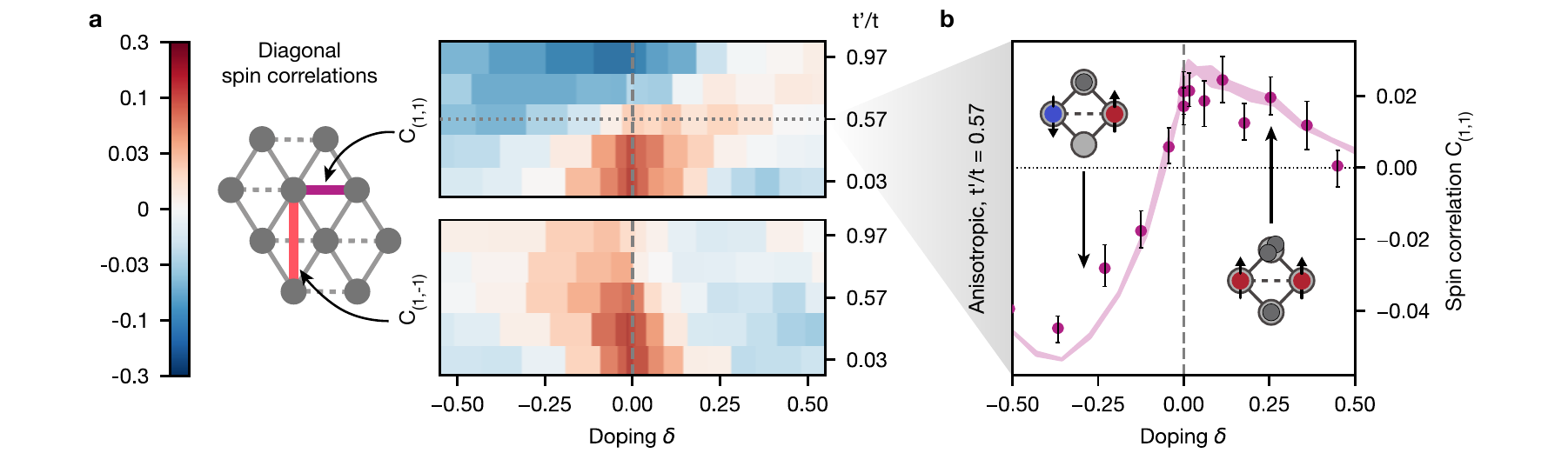"}
    \caption{\textbf{Next-nearest-neighbour spin correlations at finite doping.}
    (\textbf{a}) Spin correlations between next-nearest neighbours $C_{(1,1)}$ (along the $t'$ bond, purple) and $C_{(-1,1)}$ (diagonal, red) similarly show particle-hole asymmetry away from the square lattice geometry. (\textbf{b}) In the anisotropic triangular lattice $t'/t = 0.57(3)$, correlations along the tuneable triangular bond $t'$ show a clear sign reversal from antiferromagnetic for hole doping $\delta < -0.1$ to ferromagnetic for $\delta \geq 0$. Purple lines show a DQMC simulation at $U/t = 9$ and $T/t = 0.4$. The errorbars denote one s.e.m. and the number of repetitions can be found in Methods.} 
    \label{fig:doping_nnn}
\end{dfigure*}


%

\end{document}